# Impact of Planetary Parameters on Water Clouds Microphysics

Huanzhou Yang[1], Thaddeus D. Komacek[2], Owen B. Toon[3,4], Eric T. Wolf[3,4,5,6,7], Tyler D. Robinson[6,8,9,10], Caroline Chael[11], and Dorian S. Abbot[1]
[1] Department of the Geophysical Sciences, University of Chicago, USA; jeffyang@uchicago.edu
[2] Department of Astronomy, University of Maryland, USA
[3] Laboratory for Atmospheric and Space Physics, University of Colorado Boulder, USA
[4] Department of Atmospheric and Oceanic Sciences, University of Colorado Boulder, USA
[5] NASA GSFC Sellers Exoplanet Environments Collaboration, USA
[6] NASA NExSS Virtual Planetary Laboratory, USA
[7] Blue Marble Space Institute of Science, USA
[8] Lunar & Planetary Laboratory, University of Arizona, USA
[9] Department of Astronomy and Planetary Science, Northern Arizona University, USA
[10] Habitability, Atmospheres, and Biosignatures Laboratory, University of Arizona, USA
[11] University of Chicago, USA
*Received 2023 December 11; revised 2024 March 6; accepted 2024 March 8; published 2024 May 2*

## Abstract

Potentially habitable exoplanets are targets of great interest for the James Webb Space Telescope and upcoming mission concepts such as the Habitable Worlds Observatory. Clouds strongly affect climate and habitability, but predicting their properties is difficult. In Global Climate Models (GCMs), especially those aiming at simulating Earth, cloud microphysics is often crudely approximated by assuming that all cloud particles have a single, constant size or a prescribed size distribution and that all clouds in a grid cell are identical. For exoplanets that range over a large phase space of planetary properties, this method could result in large errors. In this work, our goal is to determine how cloud microphysics on terrestrial exoplanets, whose condensable is mainly water vapor, depend on aerosol properties and planetary parameters such as surface pressure, surface gravity, and incident stellar radiation. We use the Community Aerosol and Radiation Model for Atmospheres as a 1D microphysical model to simulate the formation and evolution of clouds including the processes of nucleation, condensation, evaporation, coagulation, and vertical transfer. In these 1D idealized experiments, we find that the parameters that determine the macrophysical thermal structure of the atmospheres, including surface pressure and stellar flux, impact cloud radiative effect (CRE) most significantly. Parameters such as gravity and number density of aerosols working as cloud condensation nuclei affect the microphysical processes of cloud formation, including activation and vertical transfer. They also have a significant, though weaker effect on CRE. This work motivates the development of more accurate GCM cloud schemes and should aid in the interpretation of future observations.

*Unified Astronomy Thesaurus concepts:* Atmospheric clouds (2180); Exoplanet atmospheres (487); Habitable planets (695)

## 1 Introduction

Thousands of exoplanets have been discovered in recent decades. The Transiting Exoplanet Survey Satellite (TESS, Ricker et al. 2010) and James Webb Space Telescope (JWST, Gardner et al. 2006), as well as ground-based surveys including MEarth, TRAPPIST, and SPECULOOS (Gillon et al. 2011; Berta et al. 2012; Irwin et al. 2014; Burdanov et al. 2022), have dramatically increased the number of detected exoplanets around M-stars. These planets, however, may not be habitable because of atmospheric loss (Deming et al. 2009; Cowan et al. 2015; Zahnle & Catling 2017). In the coming decades, missions such as the Habitable Worlds Observatory will use direct wide-field imaging at UV, optical, and near-IR wavelengths and multi-object spectroscopy to produce more detections of Earthlike planets around Sunlike stars as well as spectroscopically reveal their atmospheric composition, temperature, and potentially even biosignature gases (National Academies of Sciences, Engineering, & Medicine & others 2021).

The role of clouds remains a significant source of uncertainty in this process of discovery, strongly affecting both climate and spectral observations. Our solar system suggests that clouds and aerosols of diverse types are probably common on planets throughout the Universe. Interpreting exoplanet spectra will require a comprehensive understanding of the nature of their clouds. Past studies have found that optically thick water cloud layers obscure transmission spectra and reduce the amplitude of observed spectral features, making interpretations more challenging (Fauchez et al. 2019; Komacek et al. 2020; Suissa et al. 2020). Conversely, spatial variations in cloudiness across a planet can yield interesting and potentially useful information from reflected and thermal emission phase curves (Yang et al. 2013; Haqq-Misra et al. 2018; Wolf et al. 2019). The climate and habitability of exoplanets are also highly dependent on clouds and their radiative effects. For example, the inner edge of the habitable zone for slowly rotating planets can be extended significantly inward because of increased reflection of light to space by clouds (Yang et al. 2013, 2014; Kopparapu et al. 2016; Way et al. 2016; Kopparapu et al. 2017; Yang et al. 2019; Fauchez et al. 2021). Likewise, the choices of cloud schemes are important for the initiation of moist and runaway greenhouse states for Earthlike planets (Leconte et al. 2013; Wolf &

Toon 2015; Turbet et al. 2021). Moreover, the behavior of clouds on young terrestrial planets is critical for whether they are able to condense oceans or not (Turbet et al. 2023).

Current Global Climate Models (GCMs) used to simulate exoplanets use highly parameterized schemes for clouds, which are fast but rely on overly simplistic sub-grid-scale physics. In the microphysical cloud schemes of GCMs including CAM3 (Boville et al. 2006) and LMDZ5B (Hourdin et al. 2013), the ice particle effective radii are simple linear functions of temperature. The liquid cloud droplet effective radii are usually fixed and independent of in situ properties, although they can vary with different surface types (CAM3) or aerosol properties (LMDZ5B). ExoCAM uses a fixed effective radius (14 $\mu$m) for liquid clouds and an ice cloud effective radius that is a simple function of temperature. In ROCKE-3D and the Met Office Unified Model, liquid cloud sizes are determined by cloud condensation nuclei concentration, and ice clouds have their scattering properties parameterized with the scheme in Edwards et al. (2007). Parameterized cloud models perform well on Earth because the parameters can be adjusted to fit observational data, but this method cannot currently be applied to terrestrial exoplanets. Intra-model sensitivity tests with Exo-CAM, an Earth GCM modified so that it can be used for other terrestrial planets, found that the small changes in the water cloud particle size, aerosol–cloud interaction, and convection schemes all have a significant impact on the modeling results, with a change in global-mean temperature of $\sim$3 K (Wolf et al. 2022). With the same model, Komacek & Abbot (2019) found that different cloud particle size assumptions (from 7 to 14 $\mu$m) can change the global surface temperature of Earthlike planets by $\sim$19 K. This creates an exciting opportunity to introduce microphysical models that explicitly simulate the size distribution of cloud particles, including the effects of processes such as activation, nucleation, coalescence, and condensation on the modeling of exoplanet atmospheres.

Although a variety of cloud microphysical models have been applied to terrestrial clouds (Bardeen et al. 2013; Knutti & Sedláček 2013; Morrison et al. 2020; Archer-Nicholls et al. 2021; Proske et al. 2022), applying these techniques to exoplanets entails a number of difficulties. Terrestrial exoplanets can range broadly in their planetary parameters, including atmospheric mass, surface gravity, and stellar irradiation. Other factors including atmospheric circulation, composition, and aerosol properties also have large uncertainties. Exploring and developing models for exoplanets in this large parameter space is computationally expensive and difficult. For Earth simulations, to reduce the computational cost, models that explicitly simulate cloud microphysics usually keep some aspects of the simulation diagnostic and fixed to current climate conditions. For example, these schemes usually specify an experimentally determined criterion for conversion of cloud particles to precipitating rain or snow (Neale et al. 2010). This method has negligible impact on radiative transfer for Earth because large particles are less radiatively active, but this may not be true for exoplanets given that the cloud particle size distribution may interact with stellar irradiation differently depending on the stellar types. A fully prognostic cloud microphysical model is therefore necessary for exoplanets.

As a first step in exploring the effects of prognostic cloud microphysics on terrestrial exoplanets, in this paper we conduct a wide range of experiments using the Community Aerosol and Radiation Model for Atmospheres (CARMA), a 1D (vertical) cloud microphysical model. GCMs run column models that are linked by the atmospheric dynamics. Hence, our 1D model can eventually be transferred to a GCM. The main advantage of starting in 1D is that it allows us to explore a large parameter space in a computationally efficient way. In addition, the 1D experiments discussed here help us understand the underlying physics governing how planetary parameters affect cloud microphysics without the additional complexity, such as atmospheric circulation, present in 3D GCMs. However, it is important to remember that 1D modeling is a first step. There is a long history of using 1D models to investigate planetary cloud systems. For instance, Michelangeli et al. (1993), Colaprete et al. (1999), and Colaprete & Toon (2000) used a similar model to the one we use here to study Martian clouds with background atmospheric profiles drawn from observational data on Mars. Barth & Toon (2004, 2006) used a version of CARMA to study methane clouds on Titan. Ackerman & Marley (2001) used a 1D microphysical model, similar to the one we use, to investigate clouds of metals on hot Jupiters.

This paper is organized as follows. In Section 2, we provide a description of the models we use with the key equations. In Section 3, we describe the design of the experiments. In Section 4, we describe the baseline experiment, and we compare modeling results with observational data from Earth and Mars to address the robustness of the experimental scheme. In Section 5, we illustrate how cloud properties respond to different variables. In Section 6, we discuss the limitations of our current work and outline anticipated future work. We provide a summary in Section 7.

## 2 Model Description

### 2.1 Overview

To simulate clouds on exoplanets, we use CARMA, which originated from work by Turco et al. (1979) and Toon et al. (1988). Many of the numerical algorithms have been improved since these early papers. The microphysical model includes all important cloud microphysical processes, such as nucleation, activation, evaporation, growth, sedimentation, and coalescence. We use the model to explore how water clouds on temperate terrestrial exoplanets depend on planetary properties. Instead of tracking the bulk mass mixing ratio of cloud liquid or ice, CARMA tracks the concentration of cloud condensation nuclei (CCNs), ice nuclei (INs), and cloud particles as a function of their sizes and altitudes. In this study, we focus on simulating the processes related to the formation of water clouds from CCNs and INs on terrestrial planets. Note that CARMA has also been used to study chemical hazes and mineral clouds on hotter and larger exoplanets (Gao et al. 2018; Adams et al. 2019; Powell et al. 2019). Here we define CCNs to be water-soluble particles such as sulfates on Earth, while IN are usually not water soluble and instead have a similar crystal lattice structure to ice, such as dust on Earth. The processes that affect the concentration of particles include vertical transport due to gravity and parameterized vertical eddy diffusion ($K_{zz}$), activation of CCNs and nucleation of INs, coagulation and coalescence, growth and evaporation, and particle removal processes including precipitation.

### 2.2 Equation for Concentration Evolution

CARMA is a Eulerian cloud microphysical model that uses size bins to describe the cloud particle size distribution. A change in particle size is represented by exchange between adjacent bins. In this work, the size range is divided into 50 bins, in which the smallest radius is 10 nm, the largest radius is 800 μm, and the volume ratio between adjacent bins is 2. We make the simplifying assumption that CCNs and INs are monodisperse. This idealization allows us to more directly investigate the effects of changes in aerosol properties. Aerosols are removed by being activated or nucleated to form liquid drops or ice, and they are transported by vertical eddy diffusion and settling. However, they do not grow or coagulate. The main equation that describes the evolution of concentration as a function of particle size $r$ and height $z$ can be written as:

$$\frac{\partial C(r, z)}{\partial t} = -\frac{\partial (V_{fall} C(r, z))}{\partial z} + \frac{\partial}{\partial z}\left[\rho K_{zz} \frac{\partial}{\partial z}\left(\frac{C(r, z)}{\rho}\right)\right] + J$$
$$+ \frac{1}{2}\int_0^r \rho K_c(r', r-r')C(r, z)C(r', z)dr'$$
$$- C(r, z)\int_0^\infty K_c(r', r)C(r', z)dr'$$
$$- \frac{\partial (C(r, z) g r(r, z))}{\partial r}$$
(1)

The terms on the right side of the equation represent gravitational setting, vertical diffusive transport, nucleation ($J$), production and loss by coagulation, and growth of particles caused by the condensation and evaporation of water vapor. Instead of solving the equation directly, CARMA time-splits the equation to make it cost-effective. The time-split equation can be written as:

$$\frac{\partial C(r, z)}{\partial t} = \left[\frac{\partial C(r, z)}{\partial t}\right]_{vert} + \left[\frac{\partial C(r, z)}{\partial t}\right]_{nuc}$$
$$+ \left[\frac{\partial C(r, z)}{\partial t}\right]_{co} + \left[\frac{\partial C(r, z)}{\partial t}\right]_{grow}$$
(2)

where $C(r, z)$ denotes the particle number concentration of a specific size at a specific height, and each term on the right-hand side represents a process described above that affects the concentration. Concentrations of both ice and liquid clouds follow the same equation. Aerosols, which serve as INs and CCNs, evolve in our simulations only because of transport and precipitation removal by activation or nucleation, which are described by the first two terms. Water vapor follows a similar equation in which it is transported as well as transferred to the condensed phase through cloud particle growth, which acts as a sink for water vapor. We further describe the equations controlling each process below. Unlike most GCM cloud models, we do not distinguish rain and snow from clouds and ice as separate entities. Rather rain and snow are simply the larger particles in the size distribution that eventually precipitate out to the surface. In the model, the vertical transport and coagulation terms are solved on a longer time step, which in GCMs would be the dynamical model time step. The growth, activation, and nucleation terms are treated on a variable time step, whose length is regulated to maintain accuracy. For example, the growth time step is controlled so that growth is not allowed to switch the sign of the supersaturation in a dynamical model time step. Instead, growth proceeds slowly until equilibrium is reached.

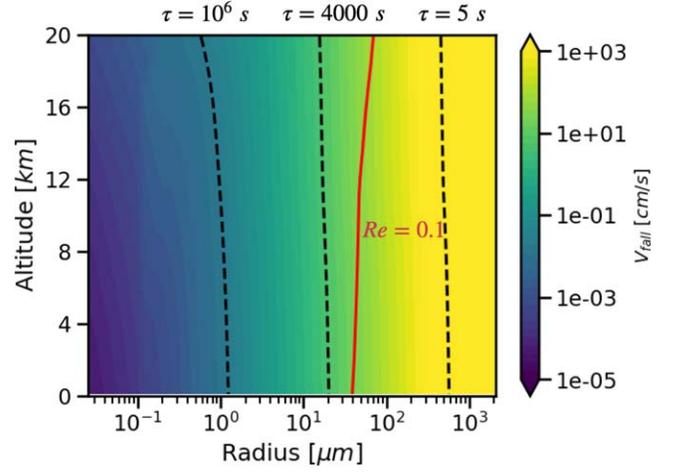

**Figure 1.** Terminal velocity of liquid droplets as a function of particle size and altitude for terrestrial conditions of gravity, temperature, and air composition. The red solid line indicates a Reynolds Number of 0.1. The black dashed lines indicate the time to fall through one of our 200 m–thick vertical levels: $\tau = 5$ s, $\tau = 4000$ s, and $\tau = 10^6$ s.

*2.2.1 Vertical Transport*

The vertical transport term is:

$$\left[\frac{\partial C(r, z)}{\partial t}\right]_{vert} = -\frac{\partial (V_{fall} C(r, z))}{\partial z}$$
$$+ \frac{\partial}{\partial z}\left[\rho K_{zz} \frac{\partial}{\partial z}\left(\frac{C(r, z)}{\rho}\right)\right]$$
(3)

The first term on the right-hand side describes the transport caused by gravitational settling, where $V_{fall}$ is the particle terminal falling velocity relative to air. The second term represents aerosol advection as a vertical diffusive process, where the parameter $K_{zz}$ is an eddy diffusion coefficient. Note that the vertical diffusion of water vapor shares the same $K_{zz}$ with cloud particles and aerosols. In GCMs, the large-scale atmospheric circulation component of this process is directly simulated, but transport due to sub-grid-scale convection is parameterized. For cloud particles, the characteristic time to reach their terminal velocity is negligible compared with the model time step (Seinfeld & Pandis 2016), so we can assume that $V_{fall}$ is always the correct fall velocity. For cloud particles with Reynolds Number $Re < 1$ (when $r < 100$ μm under a typical Earth situation), the terminal velocity is approximately $V_{fall} = \frac{D_p^2 \rho_p g C_c}{18\mu}$, where $D_p$ is the particle diameter, $\rho_p$ is the particle density, $g$ is gravity, $C_c$ is the slip correction factor (a function of $D_p$ and mean free path of air), and $\mu$ is the air viscosity. For cloud particles with Reynolds Number $Re > 1$, the terminal velocity is approximately $V_{fall} = \frac{\mu Re}{\rho D_p}$. Figure 1 shows $V_{fall}$ as a function of particle size and altitude in the control experiment, which assumes terrestrial conditions. On Earth, cloud drops are typically 5–10 μm in radius. Size classifications are somewhat arbitrary, but drops larger than 50 μm but smaller than 250 μm are usually called drizzle because they may be large enough to fall to the surface before evaporating. Typical raindrops are around 500 μm to 1 mm in radius, and rarely get larger than 4 mm (American Meteorological Society 2023). The terminal velocity is more strongly affected by the size of particles than air density in an Earthlike atmosphere. Cloud particles smaller than the dashed line on the left (falling timescale of $10^6$ s) will not fall the height of a single vertical grid box (200 m) in 10 days, the integration timescale of our simulations. Typical cloud particles will take more than a day to fall through a grid cell. Particles larger than the dashed line on the right (falling timescale of 0.5 s) are rain on Earth and have short lifetimes and small concentrations in the atmosphere controlled by the balance between particle growth rates and gravitational settling. We treat ice particles as spheres but with lower mass densities and slightly smaller terminal velocities than liquid droplets, though in reality ice particles are non-spherical and the falling velocities are more complex to compute.

$K_{zz}$ is usually determined empirically from the vertical distribution of gases and aerosols in the atmosphere. $K_{zz}$ can vary by several orders of magnitude in the vertical, mostly due to variations in atmospheric stability. In convective regions, such as the boundary layer, $K_{zz}$ is large, while in stable regions, such as the lower stratosphere, $K_{zz}$ is much lower. We vary $K_{zz}$ over a large range to represent this uncertainty. The timescale for the diffusion term is $\tau_{\text{diff}} = L^2/K_{zz}$, where $L$ is a characteristic distance. When $K_{zz} = 10^5$ cm$^2$ s$^{-1}$, $\tau_{\text{diff}} = 4000$ s (the dashed line in the middle in Figure 1) if we choose the model grid box height (200 m) as the scale length. Vertical transport of particles with larger sizes and shorter falling times than this timescale should be dominated by gravitational settling, while diffusion will be more important for smaller particles. The diffusion coefficient is not uniform in real atmospheres. For example, the boundary layer near the surface usually has stronger eddy mixing. We therefore include a 2 km layer (henceforth the mixed layer) with higher $K_{zz}$ than the rest of the atmosphere (henceforth the free atmosphere). The values of $K_{zz}$ for the boundary layer and the free atmosphere are given in Table 1. Eddy diffusion affects CCNs and INs vertical distributions more significantly than its direct effect on cloud particles, which have larger terminal falling velocity. In reality, CCNs and INs are likely transported horizontally for great distances in planetary atmospheres until they are removed by rainfall. In our model, we adopted a somewhat larger eddy diffusion than the typical values for Earth, so that the CCNs and INs would be carried upward relatively rapidly from the surface to compensate for the lack of horizontal transport from neighboring columns.

### 2.2.2 Nucleation and Activation

The nucleation and activation terms are represented by $J$, where $J$ is the rate of formation of ice particles or liquid particles, $\left[\frac{\partial C(r,z)}{\partial t}\right]_{\text{nuc}}$. $J$ varies for liquid and ice and for different formation processes. In our experiments, liquid clouds are formed from CCNs through the activation process. Activation occurs if the atmosphere is supersaturated enough that CCNs grow past the Kohler barrier. At relative humidities below 1, CCNs deliquesce and form liquid aerosols. As the humidity increases, the particles grow. If they contain enough soluble material to pass the Kohler barrier, all the CCNs will immediately become cloud droplets. The supersaturation ratio,

**Table 1**
Range of Parameters Studied[a]

| Planetary Parameter | Unit | Parameter Values |
|---|---|---|
| Incident stellar flux | Earth incident flux | 0.5, 0.63, 0.78, **1.0**, 1.1 |
| Surface gravity | Earth surface gravity | 0.71, **1**, 1.41, 20 |
| Surface pressure | Bars | 0.25, 0.5, **1.0**, 2.0, 4.0 |
| Eddy mixing diffusivity (Free atmosphere) | cm$^2$ s$^{-1}$ | $10^4$, **$10^5$**, $10^6$ |
| Eddy mixing diffusivity (Mixed layer) | cm$^2$ s$^{-1}$ | $10^5$, $10^6$, **$4 \times 10^6$**, $10^7$ |
| Atmospheric components (Viscosity and molecular weight) | / | Earthlike, $CO_2$, $H_2$ |
| CCN radius | μm | 0.05, **0.1**, 0.5, 1.0, 5.0 |
| Planet surface CCN | #/cm$^3$ | 1, 10, **100**, 1000 |
| IN radius | μm | 0.1, 0.5, **1.0**, 2.0, 10.0 |
| Planet surface IN | #/L | 0.1, 1, **10**, 100 |
| Planet surface relative humidity | % | 20, 50, 70, **95**, 99 |

**Note.**
[a] The default values of each parameter are in bold.

$S_{\text{activatecrit}}$, needed for the activation of CCNs is dependent on their soluble materials and sizes. The calculation is described in Chapter 6 of Pruppacher & Klett (2012). $S_{\text{activatecrit}}$ on Earth is usually on the order of 0.1% (Moteki et al. 2019). We assume all CCNs will immediately activate. Because our model takes time steps on the order of seconds, we have chosen $J_{\text{act}} = C(r,z)/\tau_{\text{activate}}$, where $\tau_{\text{activate}} = 0.001$ s when $S > S_{\text{activatecrit}}$. This equation is solved implicitly as a loss term for the CCNs so that only the amount of CCNs that exist can activate in a time step. For smaller values of S, $J_{\text{act}} = 0$. The number of CCNs is a critical parameter in this process. We fix the CCN number density at the surface, as listed in Table 1, so that they can diffuse upward to replace those lost by activation.

Ice clouds are formed by heterogeneous nucleation processes, which are very different from activation. INs are needed for heterogeneous nucleation to occur. This process requires a critical supersaturation to be reached. Trainer et al. (2009) experimentally determined this critical supersaturation for dust particles as a function of temperature. The contact angles of water on materials on exoplanets are not constrained. Therefore, instead of using the temperature-dependent contact angles described in Trainer et al. (2009), we assume a constant contact angle of $m = 0.95$, which is used in other terrestrial planet studies (Hartwick et al. 2019). Nucleation is probabilistic so that only a small fraction of INs are converted to ice even if critical supersaturation is exceeded. Because of the small number density of the ice particles that are typically formed, the particles can each acquire a large fraction of the available water vapor, so that their sizes become larger than the liquid drops that activate all of the CCNs and consequently have small particle sizes. The nucleation rate we use for heterogeneous nucleation is discussed in detail in Chapter 9 of Pruppacher & Klett (2012). Depending on the environmental temperature, ice particles and liquid droplets will freeze or melt. This process is described by a constant rapid conversion rate so that any particles freeze or melt in one time step. This is a simplification because we do not consider supercooled water drops (Choi et al. 2010).

### 2.2.3 Growth and Evaporation

The growth and evaporation terms are:

$$\left[\frac{\partial C(r, z)}{\partial t}\right]_{\text{gro}} = -\frac{\partial (C(r, z) R_{\text{gro}}(r, z))}{\partial r} \quad (4)$$

where $R_{\text{gro}}$ is the growth or evaporation rate of particles. The saturation vapor pressure over a flat surface varies between ice and liquid water, and is a strong function of temperature. For particles larger than the mean free path of the air, the difference between the saturation vapor pressure and the environmental vapor pressure creates a vapor gradient on the length scale of the particle size. Water vapor is transported by Brownian diffusion from or to the cloud particle. Temperature is the key parameter that regulates the rate of this process through the vapor pressure, while the supersaturation determines if growth or evaporation occurs. For liquid cloud particles, we do not consider the Kelvin effect or the solute effect on the vapor pressure, because we assume that the activated particles have overcome these barriers to growth. We are not considering supercooled water in our model, because we freeze it rapidly, so we do not allow rain to form by the Bergeron process, which is a major rain-forming process on Earth. In the Bergeron process, water vapor is rapidly transferred from supercooled drops to ice in mixed-phase clouds, creating ice particles that can be large enough to fall as snow.

### 2.2.4 Coalescence

The coalescence terms are:

$$\left[\frac{\partial C(r, z)}{\partial t}\right]_{\text{co}} = +\frac{1}{2}\int_0^r \rho K_c(r', r-r') C(r, z) C(r', z) dr' - C(r, z) \int_0^\infty K_c(r', r) C(r', z) dr' \quad (5)$$

The first term describes the production rate of particles with radius $r$ due to collisions of smaller particles, and the second term is the loss rate of particles due to collisions with other particles. $K_c$ represents the kernel for coalescence. The coalescence considered in the model is gravitational collection caused by differences in terminal velocity among cloud droplets. Like the vertical transport process, gravity and atmospheric pressure will affect this process. A critical factor in coalescence is the collection efficiency, which represents the probability that particles will actually collide and stick together to form a new particle. The efficiency is determined from the data provided in Beard & Ochs III (1984).

In this section, we described the equations used in CARMA. The equations involve variables including $V_{\text{fall}}$, $K_c$ and the nucleation rate $J$. These variables in turn depend on gravity, temperature, pressure, atmospheric composition, atmospheric dynamics (due to its effects on supersaturation), the abundance of water vapor, and the abundance and size of CCNs and INs. We summarize the values of these quantities explored in our simulations in Table 1. Because of the complex interactions of these variables, it is difficult to theoretically predict how the parameters will affect the mixing ratio and size distribution of cloud particles. We therefore use numerical experiments to analyze how the parameters affect exoplanet clouds as described below.

## 3 Experimental Setup

### 3.1 Interfacing between Models

Here we describe our scheme to simulate clouds in terrestrial exoplanet atmospheres. The scheme involves not only CARMA but also other models for initializing CARMA and post-processing of the CARMA data. We use the 1D cloud free inverse radiative convective model Clima (Kopparapu et al. 2013; Arney et al. 2016) to create the temperature profiles used to initialize CARMA. We import the temperature-pressure profile from Clima into CARMA and then keep it fixed during the cloud microphysical simulations.

Clouds affect the climate and observations of exoplanets mainly through their radiative effects. To get the cloud radiative effect (CRE), we feed the cloud data from CARMA into the Spectral Mapping Atmospheric Radiative Transfer Model (SMART, Meadows & Crisp 1996) to calculate the radiative forcing of clouds. SMART is a 1D, multiple-scattering, line-by-line radiative transfer model. It can be used for vertically inhomogeneous, nonisothermal, plane-parallel scattering, absorbing, and emitting planetary atmospheres. SMART has been used for solar system atmospheres as well as exoplanets (Robinson 2017). We calculate cloud optical depths, single scattering albedo, and scattering phase functions using a built-in Mie code in CARMA (Toon & Ackerman 1981) combined with the particle size distributions and spatial distributions of clouds. We then use SMART to post-process the cloud data for CRE. We calculate the radiative fluxes with and without clouds, and CRE is calculated from the difference between the top-of-atmosphere radiative fluxes in the two cases. Although the temperature profile varies in some cases in CARMA because of different stellar irradiation or surface pressure, we keep the temperature and moisture profile constant to that of the baseline experiment in the SMART simulations so that the resulting CRE values are easier to interpret. Only optical depth, single scattering albedo, and phase function, as well as the height of clouds, affect the CRE via changes in the planetary albedo and radiation to space. Note that some variables, including surface pressure and stellar irradiation, will significantly change the temperature profiles. We also show CRE results calculated with the corresponding temperature profiles for reference. To simulate the global-mean CRE, we use the insolation-weighted solar irradiation (510 W m$^{-2}$) scaled by the insolation values in Table 1 and a zenith angle of 49° as calculated in Cronin (2014) for Earth. Note that the 1D SMART radiative transfer calculations reflect the cloud impact in cloudy areas. The CRE we calculate will therefore be larger than the average value, including both cloudy and clear regions, because Earthlike planets should have patchy clouds that do not cover the entire sky.

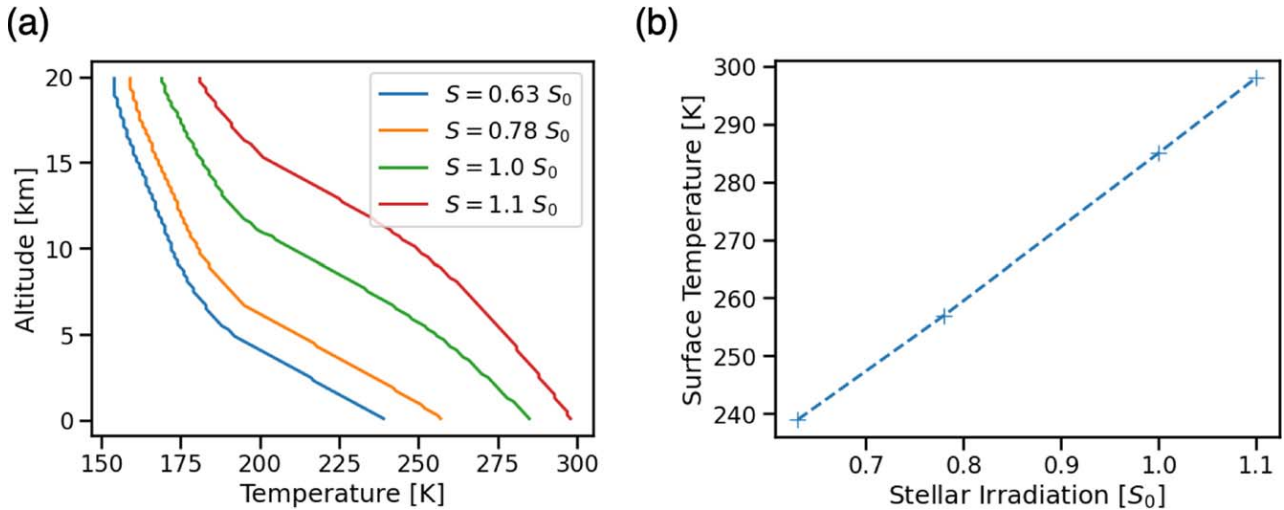

**Figure 2.** Thermal profiles from Clima, which are inputs for the CARMA simulations. (a) Temperature profiles with different stellar flux with all other parameters in Table 1 at the base case values. The green line for $S = 1.0\ S_0$ is the input for the default experiment. (b) The dependence of surface temperature with stellar flux. The surface temperature is a monotonic function of the stellar flux.

### 3.2 Numerical Details

We simulate the formation of clouds in a 1D domain with a height of 20 km and a vertical resolution of 200 m. The pressure at the model top is about 4% of the surface pressure, or 40 mbar for an Earthlike atmosphere. In test experiments, we extended the model top to 40 km and found that the cloud radiative effect changed by less than 1%, so a model top of 20 km is sufficient for our purposes. Note that transit detections can be sensitive to thinner, higher clouds as well as photochemically generated aerosols, neither of which is considered in this project.

The water vapor distribution is critical for cloud formation. In reality, supersaturation is created through a complex interplay of thermal fluctuations and cloud formation, as well as transport by convection and advection in the 3D atmosphere. In 1D we fix the temperature profile for each case and approximate the transport with vertical diffusion of water vapor. Because of the vertical temperature gradient, this diffusion process can lead to supersaturation assuming that the temperature does not increase with height. The relative humidity at the surface is fixed as the boundary condition for water vapor. Aerosols are also critical components in the formation of clouds. They can work as CCNs for liquid clouds or INs for ice clouds. The distribution of aerosols on Earth is highly dependent on their sources and atmospheric circulation, both of which cannot be properly simulated in 1D models. We fix the number densities of INs/CCNs at the boundaries. Ice clouds usually form at high altitudes, and INs that originate at high atmospheric levels, such as space dust and photochemical particles, can be important for the formation of clouds (Bardeen et al. 2008; Hartwick et al. 2019). For boundary conditions we therefore use a fixed surface number density of CCNs, and specify fixed IN number densities at both the top and bottom of the model with the same value. We set the aerosol boundary conditions to yield model interior aerosol concentrations similar to those on Earth (Wolf & Toon 2014, see next subsection). For each experiment, we assume that INs and CCNs have one uniform size so that we can isolate the effect of their size. With IN/CCN and water vapor concentrations fixed at the boundaries, we run the experiments until the cloud amount converges, usually at a scale of 10–100 days, and analyze the equilibrium state. We use a time step of 5 minutes, which tests showed was sufficient for convergence of cloud concentrations.

### 3.3 Planetary Parameter Sweep

We vary planetary parameters that are relevant for terrestrial-like exoplanets and study the effect on cloud microphysics (Table 1), including stellar irradiation, surface gravity, surface pressure, eddy mixing diffusivity, atmospheric gas species, surface humidity, CCN/IN sizes, and CCN/IN number densities. Not all these parameters are directly related to the cloud-formation equations in Section 2. Parameters like stellar insolation, which are important planetary parameters that can be constrained by astrophysical measurements, affect the vertical temperature and humidity profiles and, through them, cloud microphysics. When we calculate the initial temperature profile with Clima, we do not include clouds and we maintain a constant value for the $CO_2$ column abundance in our simulations when we change surface pressure. As a result, the greenhouse effect does not change except for pressure broadening and radiation due to water vapor, which responds to the temperature. As shown in Figure 2, with these assumptions the surface temperature is a monotonic function of the stellar flux, as is the planetary effective temperature. Besides stellar irradiation, other planetary parameters can change the atmosphere temperature profile because of their impact on radiative and heat transport. This change can be complex and introduce unnecessary noise. Therefore we change the temperature profiles only when stellar flux or surface pressure is changed. In the other experiment groups, including the group with different planetary gravities, the temperature profiles are fixed to filter out the side effects of the parameters.

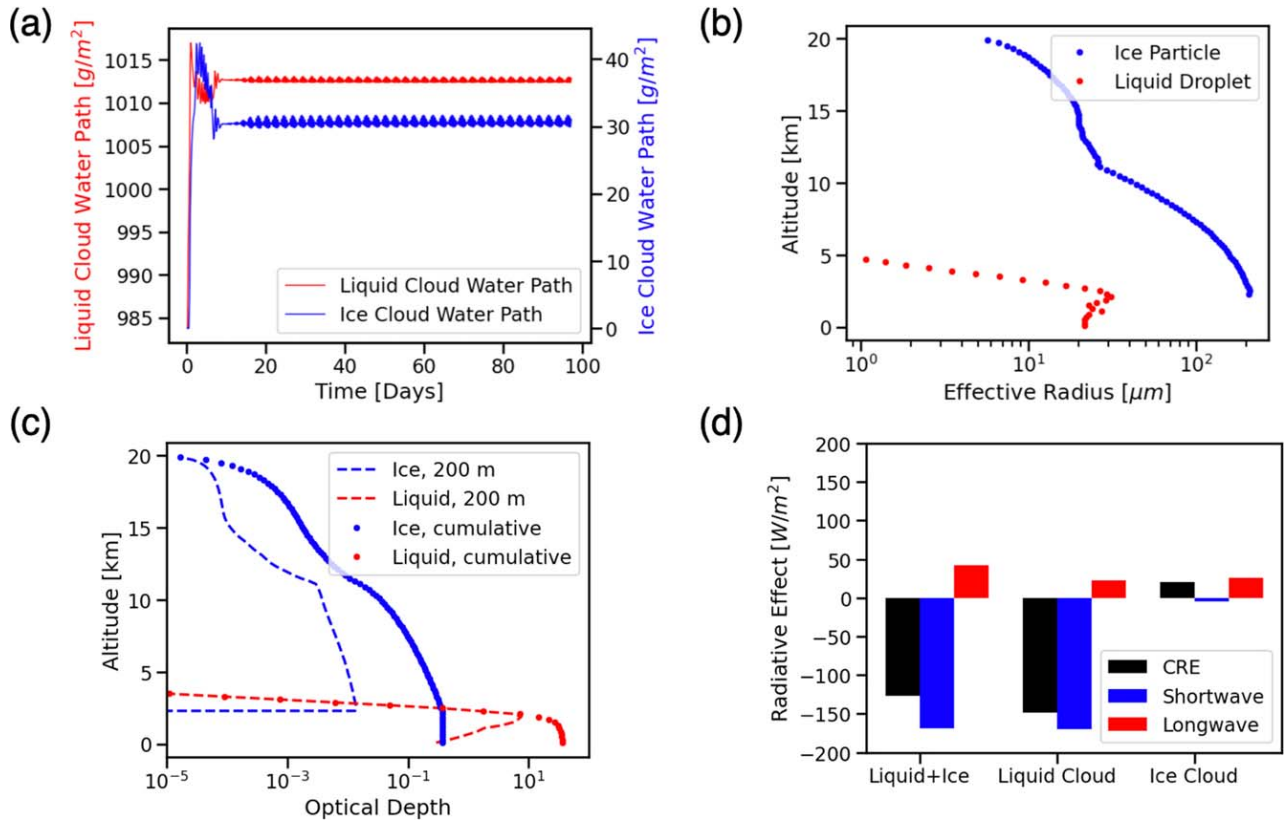

**Figure 3.** Results from the baseline experiment under Earthlike conditions. (a) Time series of ice (blue) and liquid (red) total water paths. (b) Vertical profile of ice particle (blue) and liquid droplet (red) effective radius. The effective radii for water droplets are calculated excluding particles larger than 50 $\mu$m, which are considered raindrops. (c) The cloud optical depth at 500 nm as calculated by CARMA. The blue and red dotted lines are for ice and liquid cloud cumulative optical depth from model top. The blue and red dashed lines are the ice and liquid cloud optical depth contributed by each 200 m vertical layer. (d) The cloud radiative effect calculated using SMART. Shortwave CRE includes the effect from radiation with wavelengths less than 3 $\mu$m, and longwave CRE includes the effect from radiation with wavelengths more than 3 $\mu$m. The bars marked "liquid cloud" and "ice cloud" are calculated by SMART with only the liquid or ice cloud data.

Table 1 presents the values for uncertain parameters we investigate in our simulations, with values in bold defining an Earthlike base case. Our incident stellar flux choices in Table 1 are based on the limits of the habitable zone, including cloud effects and assuming a cold start (Kopparapu et al. 2013; Yang et al. 2013; Wolf & Toon 2015). The surface gravities we choose range from a roughly Mars-size planet to the largest planet that could plausibly still be terrestrial (Rogers 2015). Similarly, we consider atmospheric surface pressure ranges relevant for terrestrial planets. We divide the column into two vertical regions with different vertical diffusivities, $K_{zz}$. The part below 2 km is the mixed layer with higher $K_{zz}$, and the higher part is the free atmosphere with lower $K_{zz}$. The depth of the mixed layer cannot be simulated well in 1D, so we fix it to a constant here, but it could depend on planetary parameters. Depending on the $K_{zz}$ value, the water vapor mixing timescale can vary from about 1 hour to about 5 days in the boundary layer. We assume that convection is moderate, with a mixing timescale for water vapor in the mixed layer of $\sim 10^4$ s, about 3 hr, in the base case experiment in Table 1, where $K_{zz}$ is $4 \times 10^6 \, \mathrm{cm^2 \, s^{-1}}$. $K_{zz}$ is a highly uncertain parameterized property for exoplanets, so the potential range of this variable in both the mixed layer and the free atmosphere is large. The size of INs/CCNs in the atmosphere has a large range and is mainly constrained by their sources and compositions. CCNs on Earth are usually water soluble, and have sizes near 0.1 $\mu$m, except for sea salt, which can be larger than 1 $\mu$m (Hudson et al. 2011). INs are usually primary particles such as dust and are larger than 1 $\mu$m (Stith et al. 2009). CCNs usually have number densities in the range of about 10–100 $\mathrm{cm^{-3}}$ over oceans and about 100–1000 $\mathrm{cm^{-3}}$ over land on Earth, while INs usually have number densities in the range from one to a few thousands per liter on Earth (Wolf & Toon 2014). We vary atmospheric gas components because their different viscosities will affect the fall velocity of cloud particles. The components we consider are Earthlike, pure $CO_2$ (Mars-like), and pure $H_2$ (low molecular mass case). The choice of gases affects only the molecular weight and viscosity of the atmosphere in CARMA, but does not affect any other properties of the atmosphere, including the temperature profile calculation in Clima or the radiative transfer in SMART. The humidities on exoplanets may also vary over a wide range. We set the surface humidity to 20%, 50%, 70%, 95%, and 99%, which represents different levels of humidity from deserts to oceanic surfaces. We set the initial relative humidity for the whole atmosphere column to the surface relative humidity value.

## 4 Baseline Experiment

In this section we will describe the clouds formed in the baseline experiment and compare the simulation results with data from Earth. The baseline experiment has parameters set to their bold values in Table 1. Figure 3 shows the cloud properties in the baseline experiment. We apply a fixed boundary concentration value for water vapor and CCNs/INs, and the simulation reaches equilibrium after about 10 days (Figure 3(a)). It should be noted that in equilibrium, the water entering the system from upward diffusion from the surface must balance the water leaving via precipitation. Hence our clouds are raining. Such a balance occurs on Earth, but on a global average rather than locally. The default diffusion coefficient $K_{zz}$ that we chose reflects global conditions and underestimates the strong vertical transport in some cloud-formation regions like deep convection, but may overestimate the transport in other locations. The simulation timescale, about 100 days, is longer than the typical timescale of individual clouds in nature. This is because cloud formation is usually a transient process, mainly due to fluctuations caused by dynamics and the diurnal cycle, which are not included in our model. We run the simulation to equilibrium because we are trying to simulate a rough proxy for global cloud behavior in a 1D model. Fluctuations around the equilibrium state in our model occur because the formation of clouds and rain have different timescales from the resupply of water vapor and INs/CCNs by diffusion. The fluctuation in cloud water paths is less than 5% of the total value, so it does not affect our main conclusions. Compared with terrestrial values, the total liquid and ice water paths are about five times larger than the largest values that occur on Earth on an annual average. We also note that clouds on Earth are only present over about 60% of the surface. Satellite observations do show individual cloud water paths as high as 1000 g m$^{-2}$ in cases like hurricanes and strong convective clouds (Meng et al. 2021; Wang et al. 2022). Because of the lack of supercooled water and Bergeron processes, this is reflected by the high liquid water path in our model.

The effective radius of cloud droplets (Figure 3(b)) is defined as $R_{\rm eff} = \frac{\int n(r) r^3}{\int n(r) r^2}$, where $n(r)$ is the number density of cloud particles with radius $r$. For liquid clouds, we excluded particles larger than 50 $\mu$m when computing the effective radius. For Earth, satellite instruments such as MODIS are not effective for particles larger than 30 $\mu$m for liquid and 60 $\mu$m for ice cloud particles (Platnick et al. 2016). For liquid clouds in our model, particles larger than 50 $\mu$m contain about 50% of the total cloud mass, but only contribute a negligible optical depth. The optical depth for mono-dispersed particles is $\tau = \frac{3}{4} \frac{Q_{\rm ext} {\rm LWP}}{\rho R_{\rm eff}}$, where LWP is the liquid water path for particles smaller than 50 $\mu$m, $\rho$ is the density of cloud particles and $R_{\rm eff}$ is the effective radius. For large particles, the extinction cross section, $Q_{\rm ext}$, is approximately 2. Therefore, for the same liquid water path, the optical depth of rain with millimeter-size droplets will be only 1% of the optical depth of clouds with 10 $\mu$m–size droplets.

We feed the cloud properties into SMART to estimate CRE (Figure 3(d)). The majority of cloud particles interact with radiation through Mie scattering. The extinction coefficient, single scattering albedo, and asymmetry factor are calculated in CARMA by the algorithm described in Toon & Ackerman (1981). The optical depth of clouds can be calculated with these properties together with the cloud mass mixing ratios from CARMA (Figure 3(c)). In this optical depth calculation, we include liquid and ice particles of all sizes.

At levels that are above freezing, mostly below 2 km, liquid clouds are dominant optically while ice clouds dominate at higher, colder levels. As shown in Figure 3(a), the total water path for ice is about 30 times lower than for liquid clouds. In addition, ice clouds typically have an effective radius that is five to 10 times larger than that of liquid clouds. Therefore, liquid clouds have optical depths that are around two orders of magnitude larger than that of ice clouds. Liquid clouds are much less extended vertically than ice clouds, so the extinction (optical depth per unit length) of liquid clouds is larger than for ice clouds, as shown in Figure 3(c). Figure 3(c) also shows the cumulative optical depth above each level. The maximum extinction of a single model level for liquid clouds occurs at about 2 km, where $\tau \sim 6$. Such thicknesses and optical depths are typical for marine boundary layer clouds on Earth. The shortwave cloud radiative forcing is mainly contributed by liquid clouds and is about $-169$ W m$^{-2}$ (Figure 3(d)). Individual ice clouds often have low optical depth, but water and ice particles are optically thick even for micron thicknesses at infrared wavelengths, hence they are excellent absorbers of infrared light. Because ice clouds are at high altitudes, where it is cold for the cases we study, they have a strong longwave cloud radiative effect comparable to liquid clouds, for a total of 42 W m$^{-2}$. The IPCC AR5 report gives the averaged shortwave CRE for Earth as $-47.3$ W m$^{-2}$ and the longwave CRE is 26.2 W m$^{-2}$. These values are 36% and 58% of the simulated shortwave and longwave cloud forcings in our model. This is reasonable given that the cloud fraction on Earth is about 67% (MODIS data, King et al. 2013) but 100% in our 1D model, and that our simulated cloud water path in Figure 2 is somewhat larger than the average value for Earth.

In Figure 4(a), we compare our simulations with satellite observations of Earth from Huang et al. (2015), which are indexed by cloud type. As our default choice of $K_{zz}$ is higher than the globally averaged value, we also have a test with a more realistic $K_{zz}$ based on data from Chapter 5 of Brasseur & Solomon (2005). In this case, the mixed layer $K_{zz}$ is $10^5$ cm$^2$ s$^{-1}$ and exponentially decreases in the free atmosphere until reaching $10^{3.5}$ cm$^2$ s$^{-1}$ at 20 km. The cloud-formation mechanism in 1D CARMA does not match exactly with any cloud type on Earth because of the lack of atmospheric dynamics and other 1D approximations. Because we set a 2 km boundary mixed layer with a higher $K_{zz}$ than the free atmosphere, the total liquid amount reaches its maximum near the top of the mixed layer, which is typical for marine boundary layer clouds. On Earth, stratus clouds above the boundary layer are mainly associated with baroclinic waves, which are not included in our 1D model. One sees these clouds in the Huang et al. (2015) data as stratus from 2 to 5 km. Despite this, the height of the liquid cloud deck in the simulations matches the observations well. The liquid cloud mass mixing ratio we simulate in the base model is 1–2 orders of magnitude larger than the global averaged value, while decreasing $K_{zz}$ makes simulation results closer to observations. This is because the humidity in our simulation (95%) is higher than the average on Earth, and represents regions where thick cloud decks form. The ice clouds in CARMA are most similar to cirrus. The simulated clouds have a similar mass mixing ratio with observations, while the height is lower. This is because the simulated liquid cloud top height is limited by the temperature where liquid droplets freeze into ice crystals, but in reality, there are mixed-phase clouds with liquid water in a supercooled state on Earth. We have not allowed such clouds in our model, and the supercooled water drop is counted as ice clouds at levels close to the liquid cloud deck. The lack of baroclinic waves in 1D mentioned above can also affect the simulation of ice clouds. More broadly, the temperature-pressure profile and dynamical features of the atmosphere vary widely around the planet such that our 1D simulations can be expected only to roughly match data. Note that the data feature a large variance, with an uncertainty of up to a factor of 2 (Huang et al. 2015).

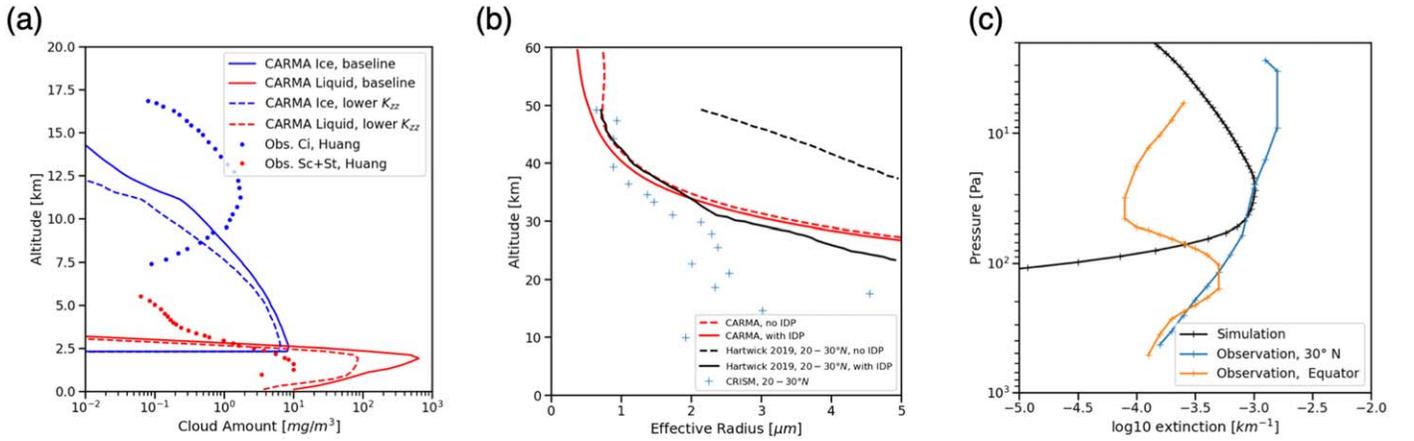

**Figure 4.** Comparison of 1D CARMA simulations to (a) observed data on Earth clouds and ((b) and (c)) simulated data from MarsCAM-CARMA, a GCM used by Hartwick et al. (2019), on Mars clouds. (a) We compare ice (blue) and liquid (red) cloud amounts from CARMA with satellite observational results. The solid lines are results from baseline experiments, and the dashed lines are results with $K_{zz}$ based on data from Brasseur & Solomon (2005). The target of comparison is cirrus for ice clouds and combined stratocumulus and stratus for liquid clouds. (b) We compare the effective radius of clouds on Mars that we simulated (red) with GCM simulation results (black) and observations from CRISM (blue crosses, Smith et al. 2013). (c) Comparison of simulated Mars cloud optical depth with observational data at the equator and 30°N. The observational data are extracted from the contour plot in Figure 1 of Hartwick et al. (2019). The observational data are extracted from published figures with the help of WebPlotDigitizer (Rohatgi 2022).

To test the versatility of our modeling framework, we also simulate cirrus clouds on Mars (Figure 4(b)). To perform this comparison, we use a temperature profile for Mars calculated by Clima. We set the upper boundary of the model to 60 km altitude to account for the fact that the majority of clouds on Mars are composed of ice and form at higher altitudes. We set the surface relative humidity to 20%, and IN concentration at the surface and model top to 300 cm$^{-3}$. This a reasonable value for dust sources from the top of the atmosphere if we consider interplanetary dust particles (IDPs) according to Plane et al. (2018). Surface dust has little impact on clouds because their height is beyond the range of transport by diffusion, so the surface IN concentration does not affect our results. We compare the simulated cloud effective radius and optical depths with observational data from CRISM (Guzewich et al. 2014; Hartwick et al. 2019) in Figures 4(b) and (c). The observations give averaged results with different latitudes, and we choose the data averaged from 20°N to 30°N for the comparison of effective sizes. For the effective radius, the 1D simulation from CARMA fits well with CRISM and a 3D GCM simulation. The ice clouds are distributed above 20 km, and the effective sizes range from 1 to 5 μm, decreasing with height in the model, as do the data. The simulated cloud optical depth is also realistic compared with the observational data cited by Hartwick et al. (2019). In Figure 4(c), we see that the simulated data are close to those extracted from Hartwick et al. (2019) at the equator and 30°N, with a maximum at around 100 Pa, and the extinction per kilometer at the scale of $\tau_{km} = 10^{-3.3}$–$10^{-2.7}$ (Figure 4(c)). IDPs in our 1D model show little impact on the ice particles, probably because we use a very large diffusion coefficient that mixes dust up from the surface. As a result, the ice nuclei from the planetary surface have a higher impact on cirrus formation than in the 3D model of Hartwick et al. (2019).

In summary, our 1D simulations provide a decent approximation of observed clouds on Earth and Mars given the limitations of our modeling framework. The goal of this work is not to represent global-mean clouds exactly, but to study the potential impact of changes in planetary parameters. The model validation we have done here supports the idea that our modeling framework is sufficient to identify the planetary parameters that have the largest effect on cloud microphysics and capture the qualitative effect of changing these parameters.

## 5 Results

### 5.1 Overview

In this section, we vary the planetary parameters listed in Table 1 and study their effect on cloud behavior relative to the baseline experiment shown in Section 4. Figure 5 summarizes the main findings of this work. The variables we vary have different dimensions. We therefore plot output as a function of the ratio of the varied parameter to its value in the baseline experiment, $\eta/\eta_0$. The range in planetary parameters reflects the plausible range in habitable terrestrial planets that allow liquid water to exist on the surface. The vertical axes are our calculated cloud properties, including CRE, cloud water path, and effective radii for ice and liquid cloud particles. The gray-shaded area shows 10% variation relative to the baseline of these cloud properties. One of our interests is the impact of clouds on climate, which depends on the radiative effect. We therefore use the cloud radiative effect as a summary statistic of the impacts of the planetary parameters (Figure 5(a)). The figure shows that surface relative humidity and stellar flux, which directly affects the surface air temperature, are the most important parameters for CRE. CCN concentration and diffusion coefficient in both the free atmosphere and the mixed layer also produce greater than 10% variation in CRE. Gravity, CCN size, IN size and number, air viscosity, and pressure variations produce less than 10% variation in CRE. We assumed the same temperature and humidity profiles in these experiments to facilitate comparison. When we change stellar flux or surface pressure, the profiles will be significantly changed. We therefore performed experiments altering the profiles for the two groups and the results are shown in Figure 5(b). Note there are two legends in Figure 5; the upper one is for Figure 5(b), and the lower one is for the rest of the figures. When we fix the profiles when calculating CRE, we underestimate the effect of surface pressure and stellar flux, but the qualitative results do not change.

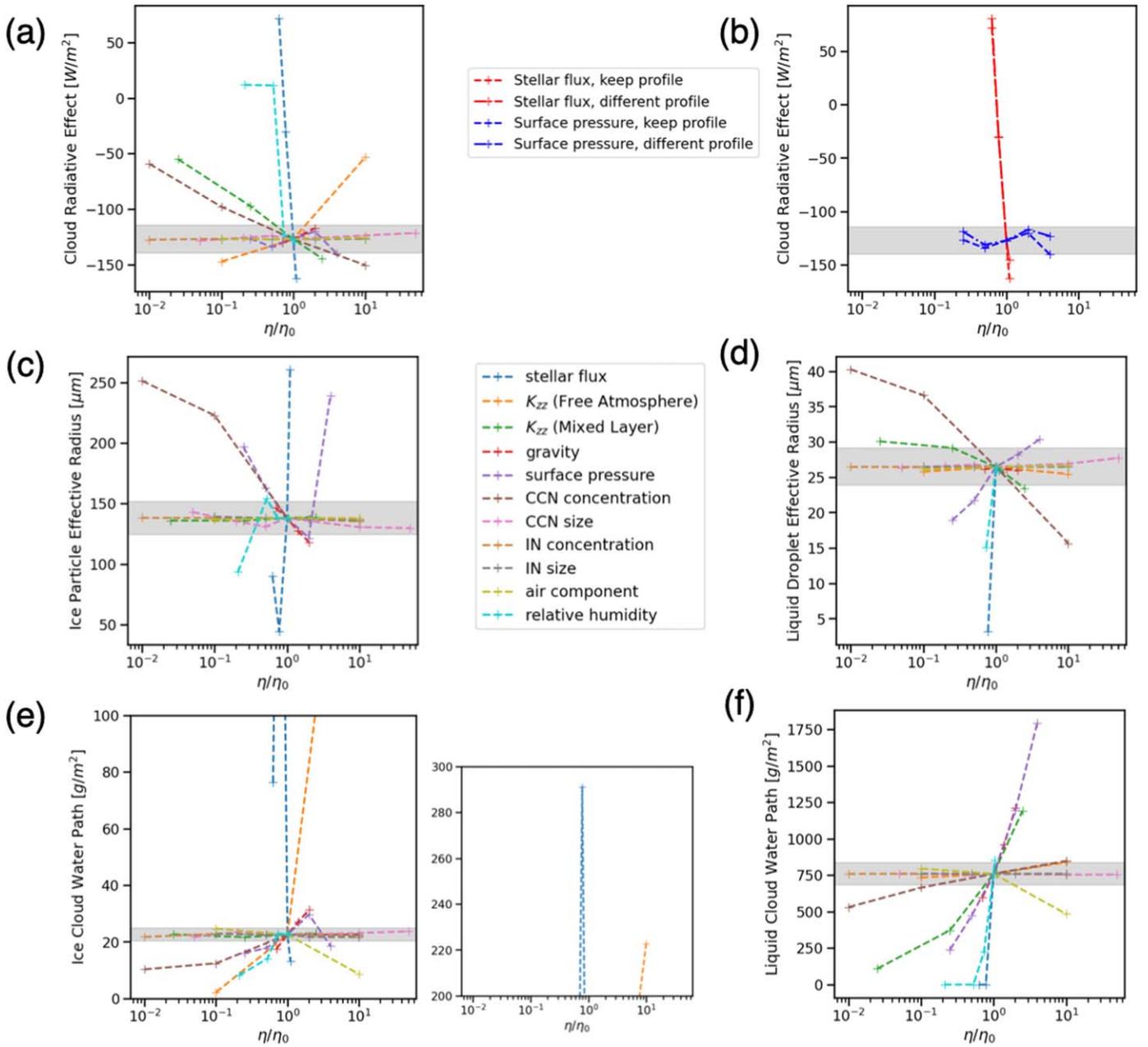

**Figure 5.** Summary of how (a) cloud radiative effects (CRE), (b) CRE calculated with corresponding $p$-$T$ profiles resulting from different surface pressures and stellar irradiations, (c) ice particles and (d) liquid droplets effective radii, (e) ice, and (f) liquid cloud water paths are affected by different parameters. The small graph near (e) shows two data points with extremely large values for ice cloud water paths. In panel (a), the CRE is calculated in SMART using the base case atmosphere profiles for all cases. The legend in the middle is for panel (a) and (c) to (f), and the upper legend is only for panel (b). The horizontal axis $\eta/\eta_0$ represents the ratio of the parameter varied to its value in the baseline experiment. For the set of experiments varying air component, $\eta/\eta_0 = 10^{-1}$ means pure $CO_2$, $\eta/\eta_0 = 10^0$ means Earthlike, and $\eta/\eta_0 = 10^1$ means pure $H_2$. The gray shade marks 10% deviation from the default value.

The optical depths of clouds are regulated by their water content and sizes, which we show in Figure 5. The effective radius is calculated as a vertically integrated quantity across the whole simulated cloud column. Though cloud sizes range over about an order of magnitude (Figure 3(b)), the mass for both ice and liquid are strongly concentrated vertically in layers where $R_{\rm eff}$ variation is limited (Figure 3(c)). Therefore this averaged value of $R_{\rm eff}$ is a useful statistic. Figure 3(d) shows that the majority of CRE is contributed by liquid clouds. It is also true that the CRE in Figure 5(a) and the liquid cloud water path in Figure 5(f) are well correlated. There is one exception in which ice clouds control CRE: when the diffusion coefficient in the free atmosphere is $10^6\,{\rm cm}^2\,{\rm s}^{-1}$, 10 times as large as in the base case (Figure 5(e)). It is the only case where the ice cloud is the main contributor to the CRE. In another outlying case, surface pressure changes result in a change larger than 10% of both effective radius and water content of both ice and liquid clouds, but the CRE change is not significant. Although liquid cloud dominates CRE when we vary the stellar flux, there is a strong nonmonotonic behavior observed in the ice cloud properties. This results from the mixed layer settings in our experiments. We will discuss the phenomena in Figure 5 in further detail in the following subsections.

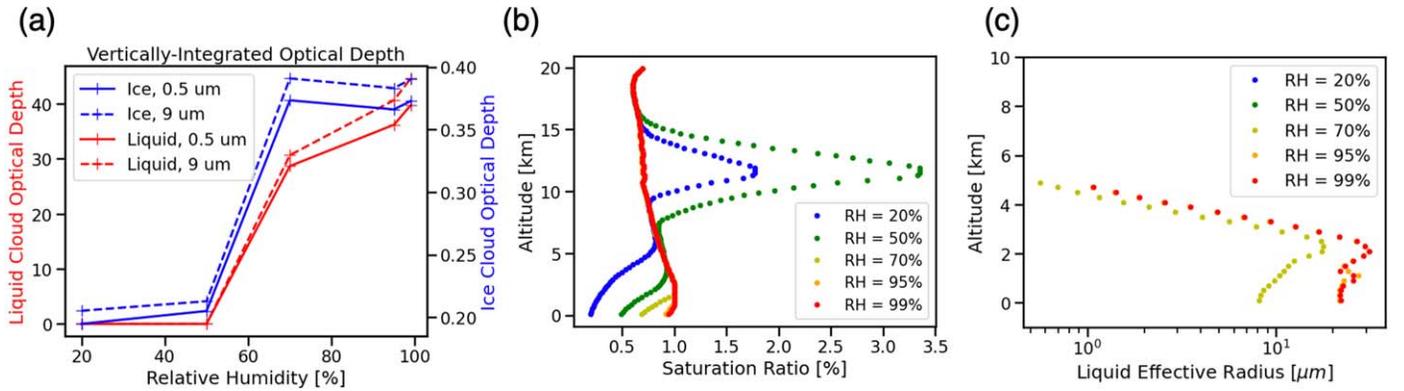

**Figure 6.** Strong response of clouds to varying surface humidity with all other parameters set to baseline values. (a) The ice (blue) and liquid (red) cloud optical depth at 0.5 $\mu$m (solid line) and 9 $\mu$m (dashed line) as a function of surface relative humidity. (b) The vertical distribution of the saturation ratio for different surface relative humidities. The temperature profile is the same for all experiments. (c) The vertical distribution of liquid droplet effective radius for different surface relative humidities. There are no liquid clouds for surface relative humidities of 20% and 50%.

We also note that variables with relatively small impact on CRE but significant impact on the effective radius should be given attention in future cloud microphysical work. Knowing the effective radius can be useful to define observational wavelengths that can penetrate the cloud. For example, if the wavelength is large enough for the cloud particles to be in the Rayleigh limit, it might be possible to see through the cloud. Weather radars on Earth choose cm-size wavelengths to detect rain and hail rather than clouds, and microwave instruments on satellites can see through clouds to measure gas properties.

### 5.2 Surface Humidity

On Earth, the huge variability of clouds is related to the different surface types, including ocean, desert, coast, and others. For exoplanets, surface humidity should also vary depending on the surface water on the planet. Aside from stellar flux, surface humidity has the strongest effect on CRE in the parameters varied in Figure 5. Surface humidity affects clouds by directly regulating the lifting condensation level, which is the altitude at which saturation occurs. Figure 6(a) shows how ice and liquid cloud optical depths are affected by surface humidity. When the surface relative humidity is 20% or 50%, liquid clouds do not exist, and ice clouds are very thin due to low abundance of water vapor. When relative humidity increases to 70% and above, which is more likely for a planet with at least partial ocean coverage, the liquid cloud optical depth increases with surface moisture. However, once liquid clouds exist, ice clouds do not change significantly with increased surface relative humidity. Figure 6(b) shows the water vapor saturation ratio with respect to liquid surfaces. The lowest point that the relative humidity becomes one represents the liquid cloud bottom. This height is determined by surface humidity. The liquid cloud top height, the place where temperature makes liquid droplets freeze, is determined by the temperature profile. The relative humidity at the liquid cloud top is constrained by saturation with liquid droplets. This value, instead of the surface humidity, directly affects ice cloud formation, which explains why ice clouds do not depend strongly on the surface relative humidity. We should still note that for clouds simulated by GCMs, the radiative feedback at the cloud top cools the atmosphere. Therefore, cirrus (ice clouds) often form as a distinct, well-separated layer from the liquid cloud deck. However, mixed-phase clouds are also common on Earth, and liquid typically composes half the cloud mass at −20°C, but mixed-phase clouds are not allowed to form in our model. We also show the effective radius of liquid clouds in Figure 6(c). When relative humidity is near saturation, there will be abundant moisture supply for the growth of liquid droplets. They can grow to particles larger than 50 $\mu$m in radius, which is defined here as rain. The cloud optical depth is inversely proportional to effective radius, and so optical depth declines for the same amount of cloud liquid water path with higher surface humidity.

### 5.3 Stellar Flux

Stellar flux helps define the temperature–pressure profile, as illustrated in Figure 2(a). The temperature profile is very important in determining the water vapor concentration, and the level at which clouds form. Liquid clouds form only at temperatures higher than 273 K in our experiment. Warmer temperatures caused by larger stellar flux lead to a large thickness and cloud water path for liquid clouds (Figure 7(a)). As shown in Figure 5, stellar flux does not affect liquid clouds significantly between $S = 1.0 \, S_0$ and $S = 1.1 \, S_0$, the two cases with a liquid cloud layer. Temperature not only affects the condensation efficiency of water vapor on cloud particles, but also affects the saturation profile, which affects where clouds form, especially with a mixed layer involved. Therefore the dependence of cloud variables on stellar flux is not monotonic, especially for ice clouds (Figure 7). When stellar flux is high enough, a layer of atmosphere near the surface will be above freezing and ice clouds will be present only above this level. When stellar flux increases and the atmosphere is warmer, ice clouds will be present only at higher altitude, so their vertical extent will be smaller, while liquid clouds will be physically thicker, and contain more liquid water because of the increase in saturation vapor pressure, increasing their optical depth (Figure 7(a)).

Temperature affects cloud properties because the amount of vapor present cannot greatly exceed the vapor pressure, which depends exponentially on temperature via the Clapeyron–Classius equation. We therefore plot the ice cloud properties with temperature as the vertical coordinate in Figures 7(b) and (c). In the free atmosphere, the ice cloud mass mixing ratio is dominated by temperature. When stellar flux is lower, the level with the same temperature will be closer to the surface, where the IN concentration is higher, because we do not have an IN source at the top of the model in these experiments. High IN concentration leads to a smaller effective radius for the same water vapor concentration.

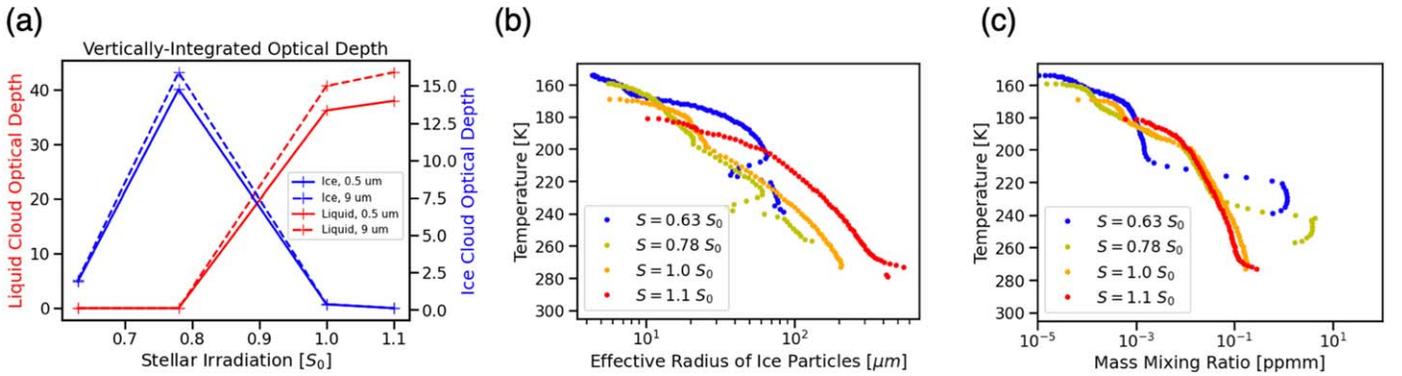

**Figure 7.** Effect of stellar flux on the following climate and cloud variables for a fixed amount of greenhouse gas: (a) ice (blue) and liquid (red) cloud optical depth at 0.5 $\mu$m (solid line) and 9 $\mu$m (dashed line). These two wavelengths are representative of shortwave and longwave radiation. (b) Effective radius of ice particles with temperature as the vertical coordinate. (c) Mass mixing ratio for ice particles with temperature as the vertical coordinate.

The optical depth of ice clouds is the highest in the 0.78 $S_0$ simulation. In this simulation, the base of the ice cloud is near the surface. Therefore the ice cloud has a large vertical extent, and has access to high IN due to proximity to the surface, which leads to high optical depth. When it is warmer (higher stellar flux than 0.78 $S_0$), the ice cloud is less extensive vertically, and has access to fewer INs, leading to lower optical depth. When it is colder (lower stellar flux than 0.63 $S_0$), there is exponentially less water vapor than in the 0.78 $S_0$ case, so the ice cloud optical depth is lower. The ice cloud optical depth reaches its maximum value when the temperature is just cold enough to allow ice clouds to form in the mixed layer, making the amount of vapor relatively high compared with the 0.63 $S_0$ case. The most significant variation in the mass mixing ratio of ice clouds is from the two coldest simulations in the mixed layer (Figure 7(c)). The existence of the thick ice cloud is determined by the high $K_{zz}$ in the mixed layer. Because of the low temperature, these ice clouds do not melt into liquid clouds, and they therefore have a high optical depth. From Figure 7(c), the cloud mass mixing ratio is larger in the two cool cases, and increases with stellar flux between the two cases. This matches well with the dependence of optical depth on stellar flux (Figure 7(a)), while the ice cloud effective radii do not show a similar dependence (Figure 7(b)). We therefore conclude that the optical features of clouds are mostly contributed by the water content, and the radius of cloud droplets is a second-order effect. Because of the existence of the mixed layer, where water supply is more abundant, the ice cloud optical depth will reach its maximum value when $S = 0.78\,S_0$ and the temperature is just cold enough to allow ice clouds to form in the mixed layer. In this scenario, the existence of abundant INs in the mixed layer is crucial. Note that INs in our simulation setup reach equilibrium with vertical distribution of gravity and diffusion, which is not always true under transient states. In cases where INs are washed out by precipitation, the conclusion may vary. However, the conclusion should remain correct for globally averaged states, because the strong vertical mixing in the lower layer generally provides abundant insoluble INs.

### 5.4 Diffusion Coefficient

Under the experimental settings described in Section 3, supersaturation is caused by the diffusion of water vapor to cold regions and to cloudy areas. The diffusion coefficient, $K_{zz}$, is the factor in 1D simulations that controls the rate of vertical transport of different species, including water vapor. The boundary layer near the planetary surface usually has rapid mixing, so we set $K_{zz}$ to a large value there. We varied $K_{zz}$ in the mixed layer and the free atmosphere separately. Figure 5 shows that varying $K_{zz}$ in both regions significantly influences CRE. Interestingly, the trends in CRE are in different directions for the two variables. Increasing $K_{zz}$ in the free atmosphere makes the cloud radiative effect less negative, while increasing $K_{zz}$ in the mixed layer makes the cloud radiative effect more negative. This is because increasing $K_{zz}$ in the free atmosphere primarily increases the mass mixing ratio of ice clouds, which have positive CRE (Figure 8(a)), whereas increasing $K_{zz}$ in the mixed layer primarily increases the mass mixing ratio of liquid clouds, which have negative CRE (Figure 8(b)), in both cases through faster supply of water vapor to the relevant region. The situation is somewhat complicated by the fact that in our model, the liquid cloud deck can extend into the free atmosphere with a warmer atmosphere, allowing an increase in $K_{zz}$ in the free atmosphere to increase both liquid and ice cloud optical depth. In reality, the depth of the mixed layer is not restricted to just the lower 2 km, but may vary.

Figures 8(c) and (d) show the different responses of ice and liquid particles to varying $K_{zz}$, which determines the moisture supply for cloud particle growth. Ice clouds usually contain large particles affected strongly by gravity. Their sizes are determined by the balance between gravitational settling and growth, in which growth is regulated by temperature. Therefore, the ice particle effective radius is not affected significantly by $K_{zz}$ in the free atmosphere. Figure 8(b) shows that as the mixed layer diffusion increases, the cloud mass increases, but Figure 8(d) shows that the effective radius decreases. Higher diffusion results in more water vapor transport up from the surface, providing more mass, but also more CCNs, resulting in a smaller droplet radius.

### 5.5 Surface Pressure

The surface pressure equals the column mass of air multiplied by gravity. The first-order effect of surface pressure on cloud formation is from its impact on the temperature profile. Clima does not take into account either the humidity we assume in CARMA or the clouds CARMA produces. Even if greenhouse gases are held constant, increased pressure leads to more pressure broadening, which warms the surface. It also leads to increased Rayleigh scattering, which raises the albedo. For the pressures considered here, pressure broadening wins so that the surface temperature increases as the surface pressure is increased (Figure 9(a)). Similar to the results when varying stellar flux, higher temperatures resulting from larger surface pressure lead to larger water cloud optical depths (Figure 9(b)) because the atmosphere contains more water vapor. The diffusion of water vapor upward from higher temperature levels increases with a larger moist lapse rate. This makes clouds even thicker with higher surface pressure.

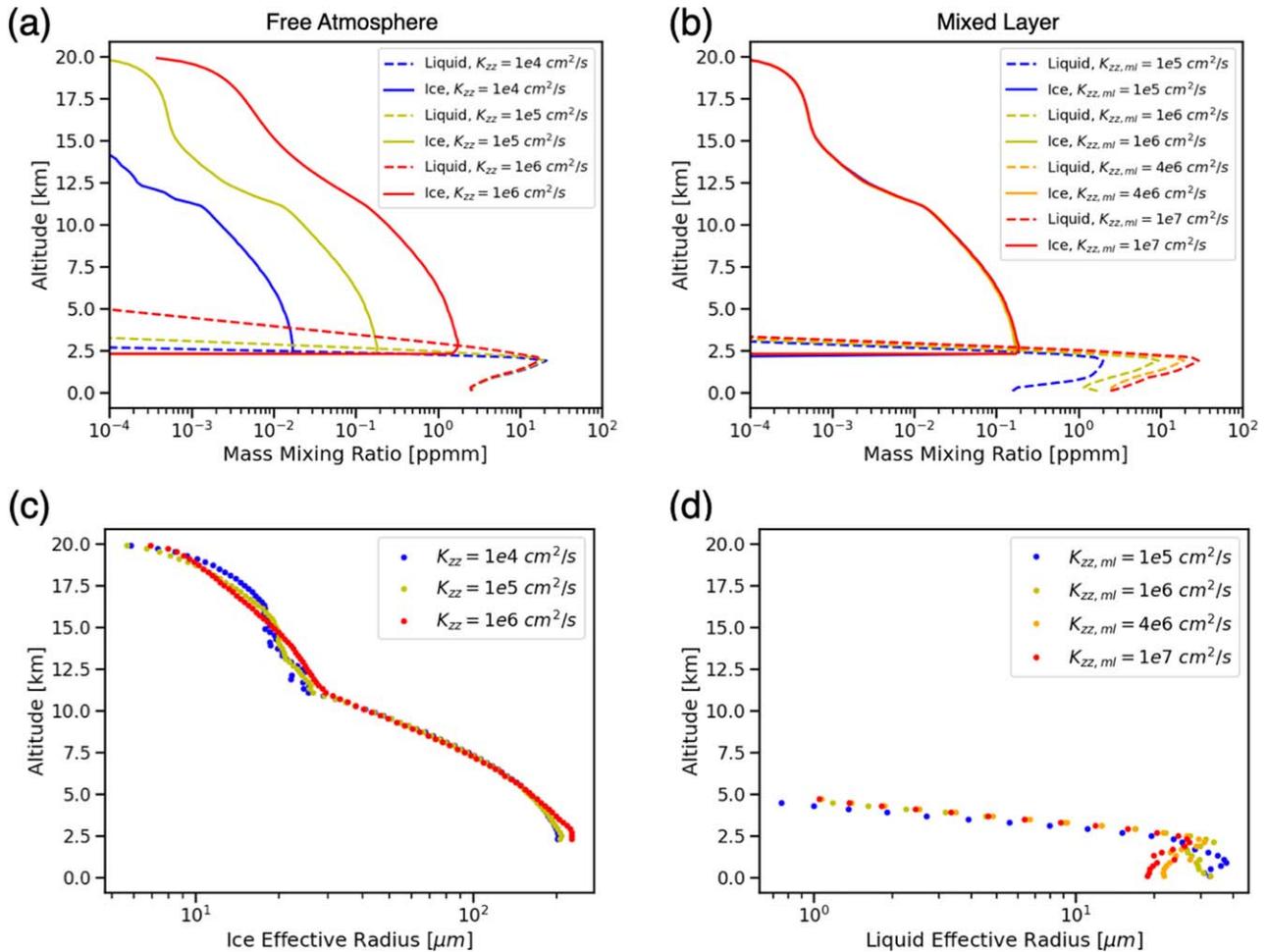

**Figure 8.** The effect of varying the diffusion coefficient. (a) and (b) Vertical distribution of ice (solid line) and liquid (dashed line) cloud mass mixing ratio when $K_{zz}$ is varied in the free atmosphere (a) and the mixed layer (b). (c) Vertical distribution of ice particle effective radius when $K_{zz}$ is varied in the free atmosphere. (d) Vertical distribution of liquid droplet effective radius when $K_{zz}$ is varied in the mixed layer.

Figure 9(c) shows the ice cloud effective radius as a function of temperature and Figure 9(d) shows the ice particle size distribution at 7.5 km. At surface pressures less than 2 Bar, temperatures are low enough that ice clouds occur in part of the mixed layer. The effective radius of ice particles is not a strong function of pressure, but more small particles are found at larger pressure as a result of more nucleation taking place because of higher temperature and water content. When surface pressure reaches 4 Bar, the mixed layer becomes too hot to allow ice clouds to form there. As a result, ice clouds form at a higher altitude, their effective radius increases, and their total optical depth drops. The number density of ice particles declines with altitude because of distance from the IN source, and to lower amounts of water in the cold upper atmosphere. Small ice cloud water path and large cloud particle size make the ice cloud optical depth decline, and hence the CRE becomes very weak at high surface pressure. Figure 5(d) shows that the effective radius of liquid clouds increases with pressure for the simple reason that it is warmer in the mixed layer, so there is more water vapor to support growth.

### 5.6 Surface Gravity

Exoplanets have different sizes and gravities. Gravity has a large impact on the atmospheric circulation and climate of exoplanets (Kaspi & Showman 2015; Komacek & Abbot 2019; Yang et al. 2019). It will affect the formation of clouds both directly and indirectly. The direct impact is that the terminal velocity of cloud particles will increase with gravity (Section 2). The indirect impact occurs through the dry adiabatic lapse rate and temperature profile as well as the column mass of water vapor. Because we have already discussed the effect of pressure and temperature on clouds when we varied the stellar flux and surface pressure, we only change gravity in CARMA for microphysical processes and not in Clima. The main effect of increasing the surface gravity on cloud microphysics is to slightly decrease the size of cloud particles, especially for ice particles (Figure 10(b)). There are few raindrops with small terminal velocity, so increasing their fall velocities slightly does not have a large impact on liquid cloud particles, which are mainly lost by evaporation or coagulation by large particles. Ice particles with larger sizes, however, are affected more strongly by gravity. Larger gravity decreases the residual time of large ice particles and results in smaller ice effective radii. Smaller cloud particles lead to larger cloud optical depth, and therefore more positive ice cloud radiative effect and less negative net cloud radiative effect (Figure 10(a)).

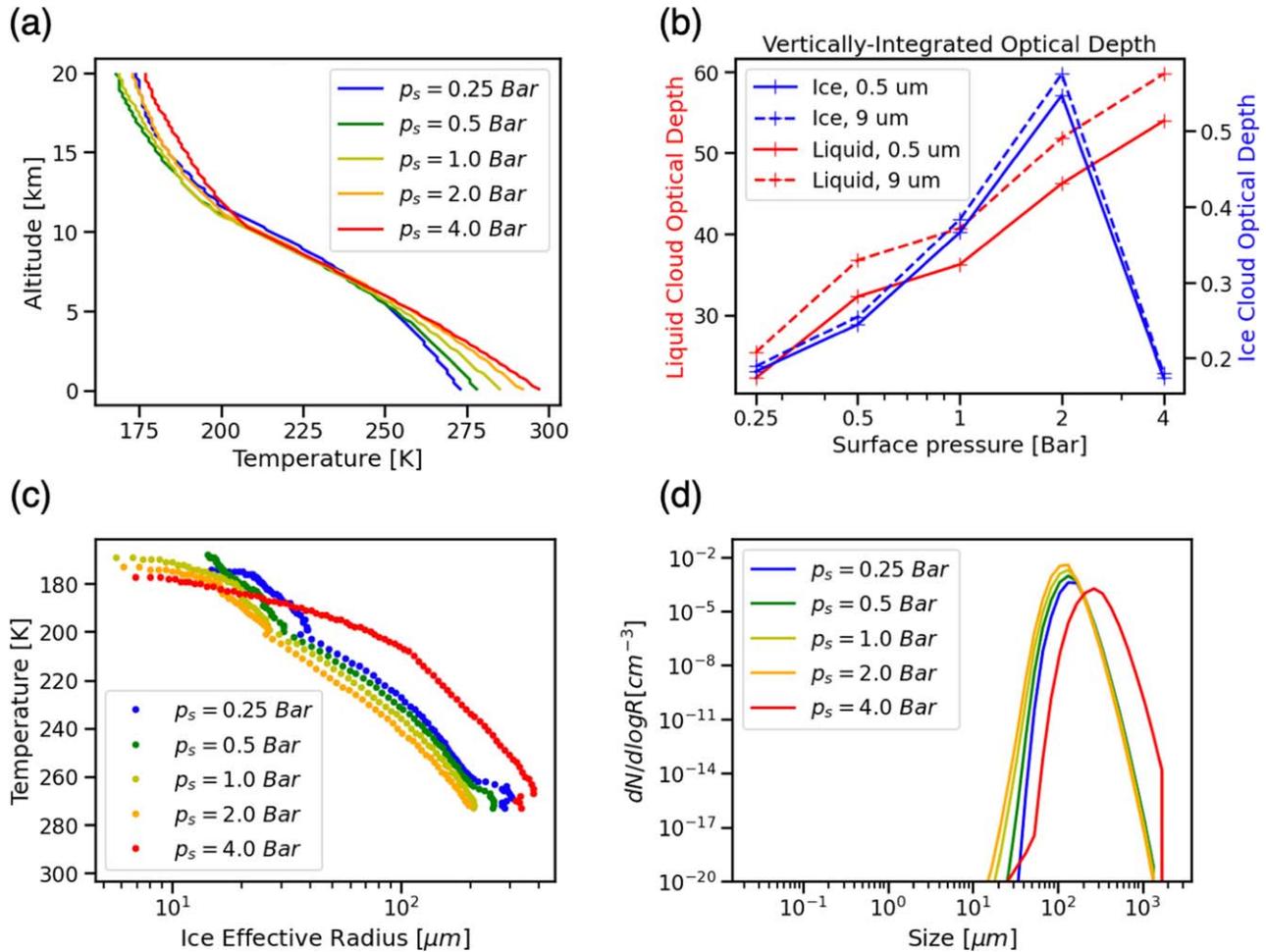

**Figure 9.** Effect of varying surface pressure on the following variables. (a) Temperature profile. (b) The ice (blue) and liquid (red) cloud optical depth at 0.5 $\mu$m (solid line) and 9 $\mu$m (dashed line) as a function of surface relative humidity. (c) Effective radius for ice clouds, and (d) distribution of ice cloud sizes at 7.5 km.

### 5.7 Aerosols in the Atmosphere

Here we consider the effect of aerosols (IN/CCN) on cloud microphysics. IN and CCN number concentrations and sizes vary widely on Earth, particularly for different source types. The impacts of INs and CCNs on cloud sizes and radiative effects are described in detail by many studies (e.g., Zhao et al. 2017; Jia et al. 2019; Barthlott et al. 2022). Our results give a semi-quantitative conclusion of how strong the impact can be on exoplanets. We find that among the sizes and concentrations of CCN/IN, only CCN concentration has a significant impact on CRE by strongly affecting the particle sizes of liquid clouds. Figure 5(d) shows that CCN concentration is the most important factor that affects the effective radii of liquid cloud particles. When CCN concentration increases over five orders of magnitude, the effective radius of liquid cloud can decline from 140 to 20 $\mu$m. With the same water content, the liquid cloud optical depth will increase by about seven times due to this decrease in effective radii. The sizes of CCNs determine the critical saturation ratio at which activation happens. However, given our model, in which saturation is forced by eddy diffusion, the impact of size on activation is small. It is well known that ice clouds on Earth do not react strongly to changes in IN concentrations. This lack of sensitivity is because nucleation occurs at high supersaturation and is stochastic. Once a small number of INs nucleate, their growth will reduce the supersaturation, preventing further nucleation. This self-limiting aspect of ice cloud formation means that adding more INs does not result in more particles forming. However, adding different types of INs with lower supersaturation thresholds has an effect, but we did not test this here, because it is likely to mainly affect 3D cloud properties, such as cloud cover. Hence the concentrations and sizes of INs have small impacts on the sizes of ice particles (Figure 5(c)), and therefore a negligible impact on the cloud radiative effect (Figure 5(a)). As mentioned in Section 5.3, our model assumes an equilibrium vertical distribution of CCNs and INs. Their specific distributions under transient states might possibly have more impacts that are not studied in this work.

### 6 Discussion

#### 6.1 Mapping Planetary Parameters to CRE

An important way we have evaluated the impact of planetary parameters is by their effect on CRE. Among the planetary parameters we varied, we found that surface relative humidity and stellar flux, which strongly affects surface air temperature, are the most important. CCN concentration and vertical diffusion coefficient also produce greater than 10% variation in CRE. Gravity, CCN particle size, IN particle size and number, air viscosity and molecular weight, and surface pressure variations produce smaller variations.

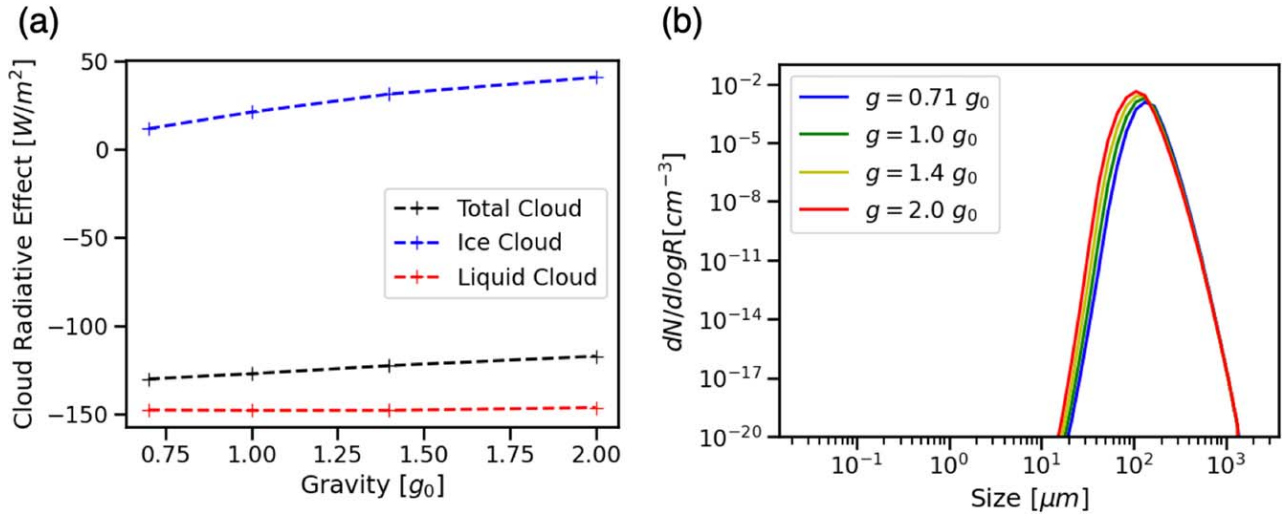

**Figure 10.** The effect of surface gravity on the following variables: (a) cloud radiative effect and (b) distribution of ice cloud sizes at 7.5 km.

Changes in cloud water content are the main driver of changes in CRE, although changes in cloud particle size have a minor impact. It is a well-established principle that the key factors regulating cloud formation include moisture, cooling, and aerosols, although the first two are strongly dependent on macrophysical processes. In our simulations, these factors regulate how planetary parameters affect CRE. Surface relative humidity and vertical diffusion coefficient determine the amount of moisture for the formation of clouds. Stellar flux and surface pressure affect the vertical temperature profile, which defines the gradient of saturation water vapor pressure. The transport of water vapor between vertical levels with different temperatures is equivalent to the cooling needed by the formation of clouds. The temperature profile influences both the cloud height and cloud water content. The existence of the mixed layer, which has substantially enhanced mixing and water vapor concentration, creates a stronger link between cloud height and water content. The formation of clouds, especially ice clouds, in the mixed layer results in quantitatively different cloud features, which we see from the group of experiments with different stellar irradiation. Aerosol properties have a relatively smaller, but robust impact on CRE. CCN number density has the largest impact because more CCNs significantly decrease liquid droplet sizes. Note that we did not consider different chemical properties and the complex size distributions for aerosols, which could also be relevant.

The parameters that do not have direct impacts on moisture, temperature, or aerosols, including gravity, atmospheric viscosity, and surface pressure, have smaller effects on CRE. They mainly affect microphysical processes after the formation of cloud particles. In particular, they affect settling and coagulation processes by changing the terminal velocities of cloud particles. These parameters do not significantly affect cloud water content. They only slightly affect cloud particle sizes (Figure 5). Despite their small impact on CRE and the habitability of exoplanets, their potential impact on observations is an important topic for future work.

*6.2 Limitations*

This work aims to provide an estimate of the magnitude of the effect of changing different planetary parameters on cloud microphysics for water clouds on terrestrial exoplanets. To perform coherent and computationally efficient experiments, we make many simplifications to the experimental settings, which are mostly related to the nature of 1D simulations. For example, we simulate the evolution of clouds until an equilibrium state is reached, whereas real clouds form and decay in transient processes that need to be studied as a time series. Moreover, our 1D eddy simulations use diffusion as a rough approximation of processes important for clouds, such as atmospheric circulation, convection, turbulent entrainment, and detrainment. When we calculate CRE with SMART, we assume a single zenith angle to represent a situation close to a global-mean state. We should note that CRE does not simply scale with stellar radiation, because planetary surface radiation does not always scale with incoming solar radiation. Considering the uneven distribution of clouds globally, a radiative transfer model coupled with 3D GCMs should calculate CRE more realistically. Some other simplifications are made for making the result representative for general situations. For example, we include only a single-size aerosol as IN or CCN. We also do not include the thermal feedback of clouds on the environmental atmosphere in order to focus on cloud microphysical processes. Although our work has limitations, it does allow us to identify the parameters that are likely to have the largest impact on cloud microphysics and understand the physical basis of their effect (Section 5) using equations in Section 2.

*6.3 Implications of Cloud Microphysics for Planetary Climate*

The 1D, horizontally homogeneous approximation is better for rapidly rotating planets than tidally locked ones, but our project may offer some insights relevant for cloud decks on tidally locked M-star planets as well. Assuming a large source of water exists, a thick cloud deck should form at the substellar point on tidally locked terrestrial planets (Yang et al. 2013).

Near the substellar point, there should be strong vertical mixing and high relative humidity, assuming the existence of an ocean. This is close to our experimental settings, which are biased toward cloudy regions rather than average global conditions for partly cloudy planets. Though a consensus has not yet been reached, it is possible that some of these slowly rotating planets have a thinner atmosphere and smaller surface pressure due to strong solar wind. Our 1D models predict larger particles for their clouds than expected from simulation results from models parameterized based on Earth. Larger particles will potentially decrease the cooling effect from the cloud deck.

### 6.4 Future Work

The formation of clouds is determined by both macrophysical and microphysical processes, and only the latter can be modeled well in 1D. Moreover, even the specific microphysical processes that dominate cloud formation differ among cloud types, which may be modulated by macrophysical processes. For example, convective cumulus clouds are dominated by the activation of CCNs, while stratospheric cirrus clouds are dominated by the nucleation of INs (Lamb & Verlinde 2011). The simulation of macrophysical processes requires modeling of 3D atmospheric circulation, waves, and convection, which is not possible in 1D. This difficulty puts importance on future 3D simulations.

Calculating the quantitative impact of planetary parameters on cloud microphysics and therefore climate will require a microphysical scheme embedded in a 3D global climate model. The simpler GCMs mentioned in Section 1 parameterize cloud particle sizes using, for example, temperature for ice clouds or surface type (which determines CCN concentration and relative humidity) for liquid clouds. This is effective for clouds on Earth; however, our results in Figure 5 show that surface pressure and stellar flux affect both ice and liquid cloud sizes significantly. Planetary parameters can therefore not be neglected for exoplanets. More sophisticated GCMs usually apply Gamma distributions for cloud particle size, which need up to four parameters fit to observations. The microphysical method used in CARMA that provides a prognostic solution with the Eulerian method is better for exoplanets, because it does not require data to fit the models as the parameterized GCMs do. Bardeen et al. (2013) simulated cirrus clouds on Earth with a GCM coupled with CARMA, which implied the computational feasibility of simulating a wider range of cloud types on exoplanets with GCMs coupled with CARMA. The 1D work in this paper provides the blueprint for parameters to focus on in 3D studies and the background theory necessary to understand the results. 3D GCMs coupled with microphysical models take a large amount of time and computational resources to run. Sweeping the parameter space we explored in this work would be too expensive in a GCM. The results in this work imply that we should put priority on simulating exoplanets with distinct thermal and moisture structures from Earth with CARMA-coupled GCMs to learn how involving cloud microphysics affects our understanding of their habitability and observational prospects.

Forward models that provide refined simulations of clouds are necessary for missions like the Habitable Worlds Observatory. Although our model with prognostic microphysics is computationally too expensive to be directly used in retrieval models, we can still employ CARMA-based models to run a large grid of experiments in advance. These can then serve as a model grid for grid-based retrieval. Additionally, the output of microphysical models can be utilized as a data set for training data-driven machine-learning models. These models run faster and are more practical for retrieval applications.

### 7 Conclusions

In this work, we ran the cloud microphysical model CARMA in 1D to determine which planetary parameters have the largest impact on clouds on terrestrial exoplanets. We simulated processes including nucleation, diffusive growth, vertical transport, and coagulation to prognostically simulate clouds. The complete configuration of our code is available online (Yang et al. 2024).

In our simulations, the parameters that determine the macrophysical thermal structure of the atmospheres are most influential for the climate and observations of temperate terrestrial exoplanets. Other planetary parameters that affect clouds through microphysical processes like gravitational settling have a smaller impact. Our work represents a preliminary analysis with a relatively low-order and computationally efficient model that allowed us to explore a broad range of parameters. This lays the groundwork to perform 3D cloud-climate simulations, including an accurate cloud microphysics scheme in a global climate model.

### 8 Acknowledgments

This work was supported by NASA award No. 80NSSC21K1718, which is part of the Habitable Worlds program. This work was supported by NASA grant No. 80NSSC21K1533, which is part of the Future Investigators in NASA Earth and Space Science and Technology program. We acknowledge the University of Chicago's Research Computing Center for their support of this work.

### ORCID iDs

Huanzhou Yang  https://orcid.org/0000-0001-8693-7053
Thaddeus D. Komacek  https://orcid.org/0000-0002-9258-5311
Owen B. Toon  https://orcid.org/0000-0002-1394-3062
Eric T. Wolf  https://orcid.org/0000-0002-7188-1648
Tyler D. Robinson  https://orcid.org/0000-0002-3196-414X
Dorian S. Abbot  https://orcid.org/0000-0001-8335-6560